\documentclass[manuscript,screen]{acmart}
\usepackage{tabularx}
\AtBeginDocument{%
  \providecommand\BibTeX{{%
    \normalfont B\kern-0.5em{\scshape i\kern-0.25em b}\kern-0.8em\TeX}}}

\setcopyright{rightsretained}

\begin{document}

\title[Tracing the History and Evolution of Dark Patterns on Twitter from 2010--2021]{Let’s Talk About Socio-Technical Angst: Tracing the History and Evolution of Dark Patterns on Twitter from 2010--2021}


\author{Ikechukwu Obi}
\email{obii@purdue.edu}
\author{Colin M. Gray}
\email{gray42@purdue.edu}
\affiliation{%
  \institution {Purdue University}
  \streetaddress{401 N Grant Street}
  \city{West Lafayette}
  \state{Indiana}
  \postcode{47907}
  \country{USA}}

\author{Shruthi Sai Chivukula}
\email{schivuku@iu.edu}
\affiliation{%
  \institution{Indiana University}
  \streetaddress{901 E 10th St Informatics West}
  \city{Bloomington}
  \state{Indiana}
  \country{USA}}

\author{Ja-Nae Duane}
\email{jduane@bentley.edu}
\affiliation{%
  \institution {Bentley University, Departments of Information and Process Management}
  \streetaddress{175 Forest St.}
  \city{Waltham}
  \state{Massachusetts}
  \country{USA}}

\author{Janna Johns}
\email{johns20@purdue.edu}
\author{Matthew Will}
\email{will10@purdue.edu}
\author{Ziqing Li}
\email{li3242@purdue.edu}
\author{Thomas Carlock}
\email{tcarloc@purdue.edu}

\affiliation{%
  \institution{Purdue University}
  \streetaddress{401 N Grant Street}
  \city{West Lafayette}
  \state{Indiana}
  \postcode{47907}
  \country{USA}
}

\renewcommand{\shortauthors}{Obi, et al.}

\begin{abstract}
Designers' use of deceptive and manipulative design practices have become increasingly ubiquitous, impacting users' ability to make choices that respect their agency and autonomy. These practices have been popularly defined through the term ``dark patterns'' which has gained attention from designers, privacy scholars, and more recently, even legal scholars and regulators.
The increased interest in the term and underpinnings of dark patterns across a range of sociotechnical practitioners intrigued us to study the evolution of the concept, to potentially speculate the future trajectory of conversations around dark patterns. In this paper, we examine the history and evolution of the Twitter discourse through \textit{\#darkpatterns} from its inception in June 2010 until April 2021, using a combination of quantitative and qualitative methods to describe how this discourse has changed over time.
We frame the evolution of this discourse as an emergent transdisciplinary conversation that connects multiple disciplinary perspectives through the shared concept of dark patterns, whereby these participants engage in a conversation marked by \textit{socio-technical angst} in order to identify and fight back against deceptive design practices.
We discuss the potential future trajectories of this discourse and opportunities for further scholarship at the intersection of design, policy, and activism.

\end{abstract}



\maketitle
{\color{red} \textbf{Draft: July 20, 2022}}
\section{Introduction}

The use of digital products is increasingly defined by manipulative and deceptive tactics. While many of these deceptive approaches are not new to digital design---instead, reflecting a translation of approaches from advertising and other fields that have long been criticized \cite{noauthor_undated-rx,Garland1964-cv}---there is a particular power that comes with the provision of services at scale that has the potential to impact users in substantial ways. The ubiquity of these deceptive patterns, often described by the moniker ``dark patterns'' is substantial. In recent scholarship, empirical work has documented that 95\% of popular apps include dark patterns \cite{Di_Geronimo2020-mh}, a plurality of end users experience manipulation as part of their everyday interactions with technology that include dark patterns~\cite{Bongard-Blanchy2021-wj,gray2021end}, and even in instances where GDPR regulation has sought to empower users to leverage their rights under data protection law, 99\% of cookie consent banners that became common in the wake of GDPR enforcement contain one or more dark patterns~\cite{Soe2020-mq}. Manipulative experiences that include the hallmarks of dark patterns and related deceptive design practices are broad and pervasive, spanning modality \cite{gunawan2021comparative}, physicality \cite{greenberg2014dark}, and context (e.g., gaming \cite{Zagal2013-gj}, privacy \cite{Grasl2021-vn,gray2021dark,Soe2020-mq}, social media \cite{Mildner2021-kp}), among others. While technology users are increasingly aware of the impact of these design practices, regulators and lawmakers are also increasingly interested in understanding and banning the most egregious practices, enforcing existing consumer protection statutes while also identifying new forms of governance to control the worst of these practices and increase user autonomy and agency (e.g., \cite{noauthor_2021-vg,noauthor_2021-bb,European_Data_Protection_Board2022-ty,Warner2021-om}).

Scholars in a range of disciplines have laid the groundwork to describe what dark patterns \textit{are} \cite{gray2018dark,brignull2015dark,bosch2016tales,Zagal2013-gj,luguri2021shining,gray2021dark}, identifying the \textit{ways in which they exist} in various technology and use contexts (e.g., \cite{greenberg2014dark,Gunawan2021-pr}), proposing \textit{comprehensive definitions} of dark patterns \cite{Mathur2021-rc}, and providing insights into how dark patterns might be \textit{detected or evaluated} \cite{mathur2019dark,gray2018dark,Stavrakakis2021-hz,Curley2021-wm}. In addition, over the last decade, technology practitioners and end users alike have used social media platforms to express their concerns about manipulative design practices, with a limited number of studies addressing \#darkpatterns on Twitter over a one year period \cite{fansher2018darkpatterns} and conversations on r/assholedesign on Reddit that relate to the use of dark patterns \cite{gray2020kind,chivukula2019nothing,Gray2021-xj}. Over the past two years, there has been increasing scholarly convergence from a range of different disciplinary and professional perspectives---including design, human-computer interaction, web measurement, and law, among others---providing insights regarding how dark patterns might be studied and effectively countered. However, there has been relatively little analysis to describe the rapid movement of dark patterns as a concept over the last decade, and the important multi-, inter-, and trans-disciplinary shifts in conversation that have informed this scholarly convergence.

In this paper, we examine the history and evolution of the the conversation around dark patterns, focusing on the Twitter discourse which has taken place through \#darkpatterns. Using a combination of quantitative and qualitative approaches, we are able to describe characteristics of this discourse, changes in participants over time, the introduction of new disciplinary perspectives, and an emerging transdisciplinary discourse that uses dark patterns as an umbrella through which many concerns relating to dark patterns can be discussed. We collected all data for this study using the Twitter API, including tweets from the first mention of \#darkpatterns in June 2010 to April 30, 2021. Through our use of quantitative and qualitative methods, including descriptive statistical analysis, network analysis, qualitative content analysis, and reflexive thematic analysis, we describe how the dark patterns discourse has evolved over its first decade on Twitter. We connect these interactions to the concept of \textit{socio-technical angst}, describing a discourse which attends to both details of technical implementation and social emergence and impact.

Through these methods, we answer the following research questions:
\begin{enumerate}
    \item In what ways has the \#darkpatterns discourse on Twitter evolved from 2010 to 2021?
    \item In what ways has the composition of participants in the \#darkpatterns discourse changed overtime?
    \item Who do participants in the discourse think are the audience of their tweets and how has this changed over time?
\end{enumerate}

This paper makes three important contributions that lay the foundation for future work on dark patterns, including identifying and curtailing deceptive design practices, at the intersection of design, HCI, law, and policy. First, we identify and characterize the ways in which the Twitter discourse related dark patterns has evolved over the years beyond its initial purpose of ``naming and shaming'' companies engaging in dark patterns, providing insight for researchers, policy makers, and technology practitioners regarding the evolution and potential trajectory of the dark patterns discourse within and beyond social media. Second, we characterize the structure of the discourse, describing its additive and increasingly pluralistic nature over the study period, surfacing how different fields adopted and engaged in the conversation from their differing disciplinary perspectives. Third, building upon a description of this discourse, we contribute insights regarding how social media users have organized to respond to unethical design practices, drawing attention to past, present, and future practices that may be useful to connect diverse stakeholders in socio-technical angst, strengthening tangible impacts.

\section{Background Work}

In this section, we situate the discourse of dark patterns on Twitter in relation to social computing and dark patterns scholarship. First, we connect our study of this discourse to the study of socio-technical activism and online social movements in Section~\ref{sec:onlinesocialmovements}. Second, we situate the specific characteristics of dark patterns discussed in this discourse within almost a decade of scholarly engagement in the topic in Section~\ref{sec:framesdarkpatterns}. Finally, in Section~\ref{sec:darkpatternsbackground} we describe the evolution of dark patterns as a concept and provide an historical backdrop to situate this evolution.

\subsection{Online Social Movements and Socio-technical Activism}
\label{sec:onlinesocialmovements}

Scholars in the CSCW community have studied online movements via numerous lenses ~\cite{savage2015participatory,kow2016mediating, stewart2017drawing,starbird2019disinformation, white2014digital,starbird2013working, lazar2019safe, gallagher2019reclaiming, starbird2019disinformation, wilson2018assembling, twyman2017black, zacklad2003communities, dimond2013hollaback}. For instance, State and Adamic ~\cite{state2015diffusion} studied how millions of Facebook users engaged in online activism in 2013 to support same-sex marriage by changing their profile pictures. Findings from their study revealed that more people participated in the campaign because they saw their friends doing so, thereby reducing the social cost of participation. Li et al. ~\cite{li2018out} designed and deployed a boycott-assisting technology to support the \#boycottNRA online social movement in their quest to boycott companies sponsoring or supporting the NRA. Findings from their research revealed that their boycott-assisting platform enabled the online activists to carry out their online campaigns in offline settings by successfully boycotting companies that support the NRA. Stewart et al. ~\cite{stewart2017drawing} also surfaced how online activists formed competing online movements to support the \#blacklivesmatter, \#alllivesmatter and \#bluelivesmatter campaigns, respectively. Findings from their research surfaced that the movements were infiltrated by misinformation agents who participated on both sides to create division and to polarize the discourse between the competing frames. Furthermore, Rho et al. ~\cite{rho2018fostering} investigated the linguistic structure employed by people commenting on the \#metoo online social movement. Results from their research revealed that the rhetorical style of respondents predicted their political leaning, and by extension, how they might respond to other related discourses. On the other hand, Dimond et al. ~\cite{dimond2013hollaback} also surfaced how storytelling plays a vital role in the sustenance of online communities. Altogether, these and other studies ~\cite{liu2017selfies,dym2019coming,rezapour2019moral,kou2017one,salehi2015we,mirbabaie2021development} foreground and characterize the diverse ways online activists have leveraged sociotechnical systems to advocate for their goals and beliefs.

Although most of the early online social movements were formed with the goal of achieving socio-political objectives, more recently, there is an emerging trend of online activism directed towards HCI/design practitioners and technology organizations ~\cite{gonzalez2019global,gray2020kind,chivukula2019nothing}, and this emerging trend is increasingly attracting the attention of regulators ~\cite{jakobi2020role}. Gonzalez et al. ~\cite{gonzalez2019global} studied public reaction to the Facebook Cambridge Analytica scandal after the company exploited user data for psychometric analysis. Findings from their study revealed that Spanish participants in the discourse placed responsibility of the scandal on the individuals (i.e., technology professionals) working at Facebook, while English users in the discourse directed much of the blame towards the company (Facebook) as an entity. Gray et al.~\cite{gray2020kind} also studied the ``r/assholedesign'' subreddit on Reddit to investigate the ways in which users called out designers and companies for the proliferation of poorly designed and antagonistic artifacts. Findings from their study revealed that users ascribed various negative qualities, which they described as ``asshole designer properties,'' based on the types of design malpractice they identified, including:
limiting user agency, information misrepresentation, ambiguity by providing conflicting information, limiting user freedom, entrapping users, and masking access requirements. Jakobi et al. ~\cite{jakobi2020role} examined the impact of regulations such as GDPR on the everyday work of practitioners, highlighting how the activities of practitioners are influenced by regulations across multiple dimensions.
We focus our attention on the Twitter discourse related to dark patterns to continue this trend of online social movements directed towards technology practitioners, focusing specifically on the nature of this discourse that includes insights from both technology practitioners and policymakers.

Furthermore, by examining the Twitter discourse related to \#darkpatterns through the lens of online social movements, we employ approaches which are native to the CSCW community to unpack and characterize both the participants, their discourse and the connections they formed to sustain the discourse for over ten years. In particular, in this paper we focus on elucidating the ways in which the dark patterns discourse has grown and evolved over the years, leveraging the characteristics of social media users that participated in the discourse, the objectives they hoped to achieve through their participation, and how these elements evolved through the discourse.

\subsection{Multifaceted Frames of Dark Patterns Scholarship}
\label{sec:framesdarkpatterns}

Numerous scholars have engaged with the increasing ubiquity of dark patterns from different disciplinary lenses ~\cite{gray2018dark,greenberg2014dark,zagal2013dark,bosch2016tales,lacey2019cuteness,mathur2019dark}. As described in more detail in the next section, the term ``dark patterns'' was originally introduced by Harry Brignull on July 28th, 2010, through the registration of the darkpatterns.org website. Harry remarked at the time that the goal of the campaign was to draw attention to the increasing use of techniques from cognitive and social psychology to manipulate users into completing a goal intended by the designers of digtial products~\cite{gray2021end,brignull2011dark,brignull2013dark,brignull2015dark}. Subsequently, Gray et al. ~\cite{gray2018dark} extended Brignull's work and defined dark patterns as the embodiment of instances when designers ``supplant user value in favor of shareholder value.'' The work of Gray et al. further identified five strategies designers employ to manipulate users into achieving their goal, including nagging, obstruction, sneaking, interface interference, and forced action. Next, Mathur et al.~\cite{mathur2019dark} analyzed around 11,000 shopping websites to investigate the proliferation of dark patterns digital shopping platforms using an automated detection process. Results from their analysis revealed that there were 1,818 instances of dark patterns on 183 of the audited websites. Based on their findings, they extended the work of Gray et al. and Brignull and included additional dark patterns of urgency, scarcity, and social proof.
Furthermore, Gray et al.~\cite{gray2021end} and Maier and Harr~\cite{Maier2020-yk} also engaged with users of technology products to explore their level of awareness, perception, and reaction to dark patterns. Findings from these studies revealed that users were broadly aware of being manipulated by digital products, and that
users distrust platforms that collect user information in an ``aggressive'' manner.
The work of Gray et al.~\cite{gray2018dark} and Mathur et al.~\cite{mathur2019dark} were later extended by Mathur and colleagues~\cite{Mathur2021-rc} to build a synthetic vocabulary of dark patterns attributes, with further guidance to describe how to consider harms relating to dark patterns and use a broader set of methods to identify and validate the presence of dark patterns.

In addition to the foundational works described above, other scholars have also addressed dark patterns from different disciplinary lenses and employing different frames. Greenberg et al.~\cite{greenberg2014dark} examined dark patterns in proxemic interactions from a critical inquiry lens.  Zagal et al.~\cite{zagal2013dark} studied the use of dark patterns in the design of games and gaming platforms and found that although most of the dark patterns they uncovered were harmless and inconspicuous, they were still problematic from an ethical perspective. Furthermore, Bosch et al.~\cite{bosch2016tales} surveyed existing literature to review why privacy-related dark patterns are effective. Findings from their research revealed that privacy-related dark patterns are effective because they employ psychological techniques to manipulate users, and the researchers created a template that users could visit to explore ways of contesting such dark patterns techniques. Lacey and Caudwell~\cite{lacey2019cuteness} also explored the use of dark patterns in the design of robots. Their objective was to draw attention to the emerging use of dark patterns in nascent technological interfaces like those found in home robots. Results from their research revealed that robot designers employ ``cuteness'' as a means of manipulating users into providing emotional data while also limiting their agency. In another context, Day and Stenler~\cite{day2020dark} investigated dark patterns from an anti-competition perspective and argued that digital manipulation techniques like dark patterns represent anticompetitive practices because they inhibit the ability of users to act rationally when making ``business decisions.'' Chromik et al. ~\cite{chromik2019dark} also investigated the potential application of dark patterns in the algorithmic explanation of intelligent systems, adding social pressure as a potential dark pattern technique that could be adopted by designers to manipulate users. More recently, Luguri and Strahilevitz~\cite{luguri2021shining} investigated user reaction to dark patterns, identifying that users reacted more vigorously to aggressive dark patterns compared to mild dark patterns. Findings from their research also revealed that the impact of mild dark patterns were more pronounced on less educated participants. Finally, Gunawan et al.~\cite{gunawan2021comparative} conducted a comparative study of dark patterns across mobile and web platforms and found that platform designers often employed different forms of dark patterns across different platform modalities, thereby further increasing the burden on users. Other scholars have also addressed dark patterns in other contexts through a diverse range of methodological lenses~\cite{di2020ui,voigt2017eu,gray2021dark,nouwens2020dark}.

Although scholars have investigated dark patterns from numerous disciplinary perspectives and in multiple contexts, limited research has engaged with the epistemic source of the conversation to investigate potential insights for technology practitioners, researchers, and policymakers. Consequently, through the lens of online social movements, we engage with elements of the discourse relating to dark patterns over a decade-long period to unpack, characterize, and surface insights into the history, evolution, and potential trajectory of the discourse on Twitter.

\subsection{A Brief Historical Timeline of Dark Patterns}
\label{sec:darkpatternsbackground}

Dark patterns are instances of unethical design practices where technology practitioners supplant the interest of users with those of shareholders, often through the use of psychological ``tricks'' or other knowledge that manipulates users into acting against their own interests~(e.g., \cite{gray2018dark,brignull2015dark,mathur2019dark,Mathur2021-rc}). The dark pattern campaign was initiated by Dr. Harry Brignull, who previously earned a Ph.D. in Cognitive Science, on July 28, 2010 through the registration of the darkpatterns.org website\footnote{This website was changed to deceptive.design in 2022 as part of a shift by some entities towards the umbrella term ``deceptive design'' due to potential racialized connotations of dark patterns claimed by some groups. This tension in naming is an active site of negotiation, with proponents of the term dark patterns describing ``dark'' as hidden or difficult to ascertain, rather than evil. In this framing, ``dark patterns'' have been contrasted with ``bright patterns'' (e.g., \cite{Grasl2021-vn,gray2021dark}), which foreground user agency in explicit ways. The impact of this recent attempt to renaming the discourse is still emerging and may become more apparent as the conversation progresses on social media.}. According to Brignull, the goal of developing the website was to employ it as a central information resource where users could report, name, and shame companies utilizing dark pattern techniques to design their products and services. The discussion around dark patterns at this time typically related to the use of deceptive or manipulative practices on eCommerce websites or related services such as airline ticket purchases. Brignull also developed different dark patterns types to help users to identify and name and shame companies employing such practices, including types such as ``sneak into basket,'' ``forced continuity,'' privacy zuckering,'' and ``confirmshaming,'' among others \cite{brignull_2010}. Following these introductory activities by Brignull and his further dissemination of the objectives of the campaign at conferences, members of the public picked up the campaign and transitioned to using \#darkpatterns on Twitter in order to publicly ``name and shame'' organizations that employed dark patterns in addition to reporting such companies to the darkpatterns.org website provided by Brignull. The discourse around \#darkpatterns has grown on Twitter based on these early developments, and the continued evolution was impacted by some key technology and regulatory events, which we will briefly summarize:

In late 2011, the BBC reported that the EU banned the use of pre-ticked website boxes as a means of protecting users from manipulation on shopping websites\footnote{https://www.bbc.com/news/world-europe-15260748}. Members of the nascent \#darkpatterns community widely shared this news and cast it as both a win for the movement and as governmental validation of their concerns. Although there was no direct connection to dark patterns in the regulation, the news of this legislation invigorated the movement near its beginning with the sense that their concerns were validated by some of the intent of the regulations. By 2015, there were multiple discrete campaigns against various forms of dark patterns in a range of contexts, including infinite scrolling in digital platforms, a campaign by designers to reign in other designers that are employing dark patterns, and celebration of a lawsuit against LinkedIn for using dark patterns.

Several notable events that impacted the \#darkpatterns discourse occurred in 2018 and 2019. First, 2018 was the year when the EU kicked off the enforcement of GDPR leading to an increase in the discourse around dark patterns from multiple disciplines including lawyers, privacy advocates, and consumer protection agencies. Also in 2018, the first broadly cited academic paper on dark patterns~\cite{gray2018dark}---and as of writing, the most cited paper in this area---characterized the nature of dark patterns, proposing a typology of five dark patterns strategies that built upon Brignull's original patterns from darkpatterns.org. This paper was the first in a now considerable body of scholarly interest in dark patterns that extends to the present day, with notable work from: Mathur et al. that describes dark patterns at scale in ecommerce settings \cite{mathur2019dark} and key definitional descriptions of dark pattern attributes \cite{Mathur2021-rc}; Soe et al.~\cite{Soe2020-mq} and Di Geronimo et al. \cite{Di_Geronimo2020-mh} that describe the ubiquity of dark patterns in consent banners and apps, respectively; and Maier and Harr~\cite{Maier2020-yk}, Gunawan et al.~\cite{gunawan2021comparative},and Gray et al.~\cite{gray2021end} that indicates the ubiquity of manipulation in users' digital experiences in conjunction with a lack of knowledge on how to effectively respond. Finally,
2018 was also the year when the Norwegian Consumer Protection Council produced a report on the dark pattern practices by Google and Facebook---the beginning of a trend of multiple consumer protection and regulatory bodies in Europe leveraging dark patterns to describe consumer harms and create the foundation for sanctions.

From 2021 onward, the regulatory interest around dark patterns became solidified in parts of North America as well, with the US Federal Trade Commission conducting a workshop and exploring ways to regulate deceptive practices inscribed into dark patterns. The term ``dark patterns'' is now commonly used in both legislation---for instance in the proposed DETOUR Act in the United States~\cite{Warner2021-om}, the recently passed Digital Services Act in the European Union~\cite{noauthor_undated-qf}, and CPRA legislation in the State of California~\cite{noauthor_undated-ae}.

\section{Method}
In this study, we began by conducting a multi-year digital ethnography \cite{Boellstorff2012-vn} from 2017 to 2021 in order to situate our research engagement and sensitize ourselves towards features of the \#darkpatterns discourse on Twitter. First, we began our investigation by following relevant accounts (e.g., \@darkpatterns) that were identified from previous research by Fansher et al.  \cite{fansher2018darkpatterns} and collected common hashtags associated with the topic.
Following our exploratory investigation and extended researcher sensitization to the discourse, we focused on the \#darkpatterns and \#darkpattern hashtags to observe and collect salient tweets, building upon related work (i.e., \cite{fansher2018darkpatterns,gray2018dark}). Next, after observing the discourse from 2017 to 2021, we employed our selected hashtags to retrieve all historical tweets relating to dark patterns within our date range from the first ever use of the hashtag in June 2010 to April 30th, 2021 via the Academic Twitter API v2. In total 30,951 tweets were collected for analysis.

We approached our characterization of the use of \#darkpatterns on Twitter as a multi-dimensional conversation, and following best practices of previous research \cite{starbird2019disinformation,arif2018acting}, utilizing multiple analysis methods to unpack, connect, and make sense of the discourse. To this end, our analysis pairs quantitative approaches such as descriptive statistics and network analysis with a qualitative content analysis approach to examine and describe the evolution and potential trajectory of this discourse.

\subsection{Data Collection}
We collected data for this study via the Twitter API v2 for Academic researchers. We captured tweets that included the hashtags "\#darkpatterns" or "\#darkpattern", and "\#dark \#patterns" in their content starting from the first-ever use of the hashtag \#darkpatterns in 2010 until our data collection cutoff date of April 30th, 2021. Across this time range, we captured 30,951 tweets which were then stored in a relational database for later analysis. This initial dataset included accompanying data and metadata about tweet each including the following; tweet\_id, author\_id, tweet\_text, lang, created\_at, hashtags, retweet\_count, reply\_count, quote\_count, like\_count, and referenced\_tweets, among other data points. Following our first round of data collection and by using the author\_id attached to each tweet, we conducted secondary data collection to retrieve the profile information of the users that participated in the discourse. Since we are interested in characterizing the affiliation of participants in the discourse, our data collection for this second round focused on the user name and user description, among other data points, including following count, follower count of each of the participants. We captured this information on the 17,993 unique Twitter accounts that used the variants of the dark patterns hashtag in their tweets within the specified time range.

\subsection{Data Analysis}
We employed complementary quantitative and qualitative data analysis approaches to answer our research questions which included: 1) descriptive statistics, hashtag analysis, content analysis, and reflexive analysis approaches to answer RQ \#1, which explores the ways in which the Twitter discourse related to \#darkpatterns has evolved from 2010 to 2021, 2) network and content analysis to answer RQ\#2, which seeks to investigate the characteristics of participants in the discourse, and 3) qualitative content analysis to answer RQ\#3, which examines the intended audience of tweets generated in the discourse. We describe these approaches in detail in the following subsections.

\subsubsection{Descriptive Statistics}
We used descriptive statistics \cite{holcomb2016fundamentals} to describe how the Twitter discourse related to \#darkpatterns has evolved from 2010 to 2021. Our analysis focused on analyzing data regarding the number of tweets and retweets for each year, the number of unique Twitter accounts that participated in the discourse each year, and the different languages represented in each year of the discourse. We then compared these data points by year, demonstrating that user concern about dark patterns on Twitter has increased over time and has---almost since its origin---cut across multiple languages in multiple parts of the world.
We also described common hashtags that co-occurred with the variants of the dark patterns hashtags and how the hashtags have changed as the discourse progressed. Tweet data were processed through a script to identify a count of the hashtags collected from the discourse for each year and across all the years of the discourse, followed by a reflexive qualitative curation of technology related hashtags by the researcher team. Prior research ~\cite{rho2019hashtag} has shown that social media users use hashtags to make their content discoverable and also to connect other domains (represented as hashtags) to the conversation, and we use this lens to identify the domains and topics that were most frequently connected to the dark patterns discourse for each year of the data we collected. We further analysed the most shared link resources in each of the discourse to examine the ways in which they expose the shared concerns around \#darkpatterns during each year of the discourse.

\subsubsection{Network Analysis} \label{networkanalysis}
We employed network analysis to identify dominant professional fields represented in the dark patterns discourse and the interactions between participants represented by these fields. We define interaction in this context as a retweet, quoted tweet, or reply to a tweet about dark patterns.
In constructing this network, we used Twitter accounts as nodes and retweets, replies, and quoted tweets as the edge connecting two nodes. The relative size of the node is determined by the number of retweets, shares, and quoted tweets linked to that node. For interpreting the network graphs, we related the nodes to the professional fields they represented based on their Twitter user description in place of using Twitter accounts or usernames.

This network analysis was conducted via an interpretivist analysis process \cite{starbird2019disinformation,arif2018acting} which allowed the researchers to continuously with discuss each other, the potential stories that could be framed by these visualizations. We first created a network graph for each year of the discourse. Based on the findings from this analysis process and through discussions, we merged relatively similar network graphs into three phases that best characterizes the evolution of participation within the discourse. These phases included: 2010 to 2015 (The Era of Intra-Disciplinary Discourse), 2016 to 2018 (The Rise of Legal and Regulatory Interest), and 2019 to 2021 (A Concretization of Transdisciplinary Discourse). These phases further aligned with the periods before the announcement of GDPR, the announcement of GDPR, and the early enforcement of GDPR. Thus, building this network allowed us to surface the primary accounts and related professional fields represented during the various phases of GDPR adoption and enforcement interest. We employed the Yifan Hu algorithm ~\cite{hu2005efficient} to build the the network graph layout due to its suitability in segmenting the subgroups within our network in a clear way. We also used the Louvain algorithm ~\cite{blondel2008fast} to surface the communities within the discourse.

\subsubsection{Content Analysis}
We employed a qualitative content analysis technique inspired by Neuendorf~\cite{neuendorf2017content} and Gray et al.~\cite{gray2020kind} alongside a reflexive thematic analysis approach ~\cite{braun2019reflecting,Braun2021-dt} to answer RQ \#3. As a research team, we conducted a qualitative content analysis of 1) all tweets from 2010 to 2014, 2) all the unique retweets in the discourse, and 3) randomly sampled tweets from 2015 to 2021. Tweets and retweets were coded separately. We coded all tweets from 2010 to 2014 to gain foundational insight into the early evolution of the discourse. We coded the unique retweets (all original tweets that were later retweeted) across the entire dataset to explore if they share attributes or commonalities. Next, we coded a random sample of tweets from 2015 to 2021 within the dark patterns discourse that represented approximately 15\% of the total number of tweets for each year. Altogether, we coded 5,830 tweets and the results from this analysis provided insights that allowed us to address research questions 2 and 3.

Eight researchers participated in these coding sessions over a six month period, part of a research team from three different US institutions. The majority of the research team was located at a large research-intensive public institution, including the primary investigator, graduate researchers, and undergraduate researchers.

All researchers, including three graduate and three undergraduate students had previous training in qualitative research and had experience conducting qualitative content analysis in previous research projects. In all, our content and thematic analysis process followed best practices to increase trustworthiness and rigor \cite{Maxwell2004-fr}, including sensitization with the data set, open coding, reflexive engagement with a codebook to concretize our coding approach, confirmatory pair coding, and construction of meaningful themes through a reflexive process.
We describe our qualitative analysis approach for each portion of the dataset below.

\paragraph{1. Getting familiar with the data set}
We commenced the analysis process by conducting a preliminary coding of the data set to reflect on a) who appears to be posting the tweet, 2) who their tweet appears to be directed towards, and 3) what kind of issues or concerns were expressed in the tweets. In total, we open coded 254 tweets randomly selected using a random number generator from the larger dataset to sensitize ourselves to the discourse and guide preliminary directions for further analysis. Based on the review of the codes from our preliminary analysis, we then developed a codebook to guide the next coding session that yielded the data set for this study. These coding sessions were supported by a custom web-based coding interface developed in-house by our lab for conducting qualitative content analysis that connected to a read-only version of the relational data captured from the Twitter API.

\paragraph{2. Creating, adjusting, and validating the coding scheme}
During the second round of coding we utilized the code book that was developed from the preliminary analysis. We separated the data set according to tweets and retweets and coded them separately since they had a different function in the discourse. We coded all the unique tweets that were retweeted in the data set to explore if there are common themes or threads that are shared. We also coded all of the tweets from 2010 to 2014 to gain a foundational understanding of the early stages of the discourse as another form of data analysis and sensitization. Next, we then coded approximately 15\% of tweets for each year from 2015 to 2021, using a random number generator to create a manageable, while still representative, sample from the dataset.

In this phase of analysis, our final codes listed in Table ~\ref{codebook} included four primary categories, informed by our prior open codes and sensitization to the dataset : 1) what the tweet was about; 2) who appeared to be posting the tweet based on the profile bio or tweet text; 3) who the post appeared to be directed towards, and 4) what type of issue was expressed in the post. All codes were applied non-exclusively and open text boxes were used to identify further nuance that we addressed in our later reflexive analysis.

During these coding sessions, we continuously engaged in conversations about how well the codebook was suited to our analysis, collectively deciding to refine the codebook or identify different inclusion and exclusion criteria for specific parts of the content analysis where necessary. The researchers were also encouraged to create memos on tweets which they considered important and relevant to the research questions for further group discussion. At the end of this round of coding activities, we conducted a pair-coding exercise between two coders to review all applied codes to reach a consensus on the application of codes and also ensure that all the codes were applied according to the guidelines in the codebook.

\paragraph{3. Reflexive Analysis}
During this final round of coding, we conducted a reflexive thematic analysis ~\cite{braun2019reflecting,Braun2021-dt} by focusing on the tweets that different members of the team had memoed as interesting or otherwise marked for further inspection during the earlier coding sessions. In total, we analyzed 404 tweets that were distributed across all twelve years of the dataset. We used Miro, a digital whiteboard tool, to individually organize these tweets as digital Post-It notes, using metadata tags to indicate which year each tweet was posted.
Four researchers participated in this reflexive analysis, with each researcher coding the same dataset independently and generating codes and and potential themes that they felt best characterized aspects of the discourse. Following this individual analysis, through extensive conversation as a research team, we constructed final themes to characterize key aspects of the discourse that supplemented our other forms of data analysis.

\paragraph{4. Analyzing User Descriptions}
We adopted a somewhat different analysis approach to investigate and characterize the user descriptions of participants in the coded tweets. We began by leveraging the open codes provided by the researchers through our initial analysis, which resulted in 733 distinct professions or roles based on the profile description of the Twitter user. We then discussed various combinations of user role groups based on the open codes, resulting in 18 distinct categories of roles that retained the diversity implied by our initial open codes in a meaningful way.
For example, UX Designers, Graphic Designers, and Product Designers were all categorized as Designers. Similarly, we sought to ensure that the diversity of roles was also retained, with distinct role categories for Software Engineer/Developers, Data Privacy and Security Specialists, and upper management, for instance. We also sought to characterize different groupings beyond technological professions, for instance, creating categories for professional organizations, entities relating to public policy and consumer rights, legal professionals, journalists, regulatory agencies, and academics. The complete list of these categories is summarized in Figure~\ref{fig:userroles}. Based on our final categories, we exclusively assigned each open-coded tweet with one of our final role categories.

\begin{table*}[!htb]
    \begin{tabularx}{\textwidth}{p{.35\textwidth}X}
       \toprule
       \textbf{Code} & \textbf{Explanation of code and options } \\
       \midrule
1. \textit{What is this tweet about?} & \textbf{A. Example or Experience:} sharing an example or experience with dark patterns, \textbf{B. Meta:} talking about the fact that there is a discourse but not engaging in the discourse, \textbf{C.News Article:} talking about or sharing news article about dark patterns, \textbf{D. I don't Know:} won't see the rest of the interface\\

2. \textit{Who is posting the tweet?} & What is their Twitter account user description? \\

3. \textit{Who is this post directed towards?} & Who does the user think they are talking to? \textbf{A. Organization:} calling out an organization, \textbf{B. Government:} calling for action/attention towards legislation, \textbf{C. Professional Group or Affiliation:} Calling out or mentioning a particular group of professional roles, \textbf{D. Personal Network:} Personal experiences and examples | talking to a particular person | “Hey my peeps”, \textbf{E. Unidentifiable}: I don't know. \\

4. \textit{What type of issue is being expressed in the post?} & What is the implied discourse around dark patterns happening in this post? \textbf{A. Designers, Product UX and responsibility}, \textbf{B. Regulation around dark patterns}, \textbf{C. Research around dark patterns:} calling for research, need for more research, connecting research of above two (product UX and legislation), sharing research, \textbf{D. News coverage around dark patterns:} sharing news article about dark patterns. And other mass media coverage of dark patterns\\
       \bottomrule
     \end{tabularx}
     \caption{Codebook for the content analysis of the dark patterns dataset. All options used in the coding interface were applied non-exclusively.}
     \label{codebook}
 \end{table*}

\subsection{Research Ethics and Positionality}
Studying social media data necessitates ethical considerations and sensitization ~\cite{fiesler2018participant}. We collected data for this study via the Twitter API v2 for Academic Researchers and ensured that our work complied with the requirements and terms of service of the platform. Researchers that participated in the data collection process sensitized themselves to the ethics of conducting research with social media data, for instance, by reading and discussing the ethical guidelines adopted by AoIR \cite{Franzke2020-ha}. Furthermore, although the content of most of the tweets is public, the owners of the Twitter accounts did not directly consent that we use their tweets for research. Hence, we sought to limit the personal detail we included in our analysis to public user descriptions, and all tweets from the dataset that we describe in this paper are paraphrased to reduce discoverability of the original tweet. Additionally, through our analysis of user roles, we have sought to protect the anonymity of the users by collapsing their specific role into a smaller number of categories. As important exceptions to these ethical standards, we do reference a limited number of examples of tweets or user accounts that are specifically indicated as organizations or participants within the public interest (e.g., regulatory organizations such as \@FTC and \@Forbrukerradet; lawmakers such as \@MarkWarner) or are intended to promote activism in a public way (e.g., \@darkpatterns, \@harrybr). In general, these small number of more public accounts have a high number of followers and tweet with the intention of broad propagation and engagement, which we feel is in the spirit of our goal of building awareness of ethically-problematic design issues.

In addition, we acknowledge our areas of disciplinary expertise which has impacted the kinds of analysis we were able to conduct, the themes and codebook which we constructed, and the framing of our findings. Our research team includes members trained in computing, design, psychology, and ethics and all members care deeply about increasing ethical consciousness in technology and design practice.
We also build upon our prior interest and research on issues relating to technology ethics and dark patterns, which have sensitized us towards particular aspects of the discourse as it has evolved over time.

\section{Findings}

\begin{figure}
    \centering
    \includegraphics[width=\textwidth]{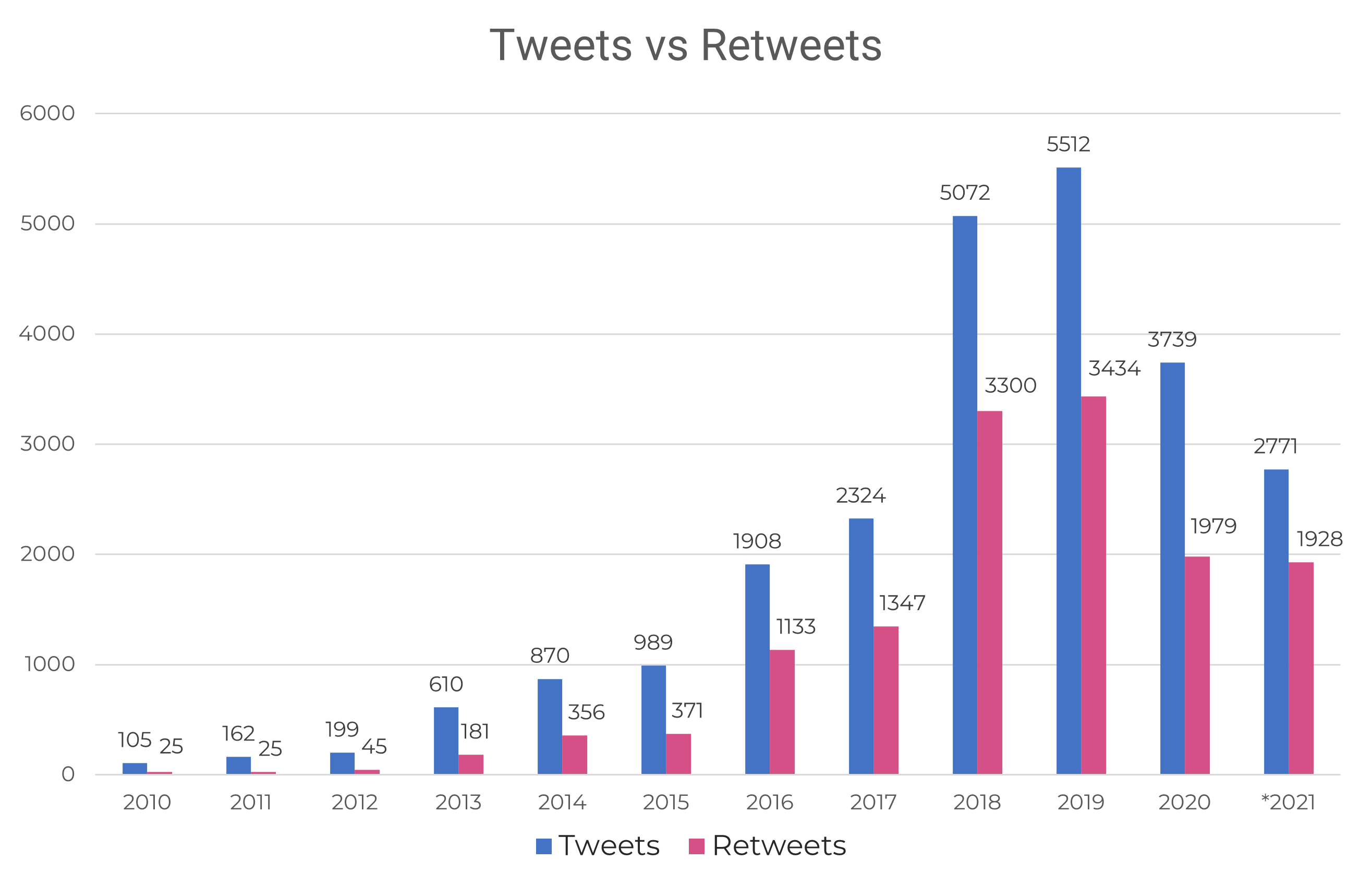}
    \caption{Tweets and retweets for each year of the discourse (from June 2010 to April 2021).}
    \label{fig:tweetsvsretweets}
\end{figure}

\subsection{RQ1: In what ways has the \#darkpatterns discourse on Twitter evolved from 2010 to 2021?}

To answer RQ \#1, we employed used a combination of descriptive statistical analysis, content analysis, and hashtag analysis to identify how the discourse has evolved over an ten year period (2010 to April 30th, 2021). Descriptive statistical analysis \cite{holcomb2016fundamentals} illustrates how the discourse has expanded in terms of the number of participants and tweets, showing an increase of unique Twitter users in the discourse from 95 in 2010 to 3449 in 2020, with a peak of 5425 in 2019. Over the ten year time period, the number of tweets increased from 110 in 2010 to 5032 in 2020, with a peak of 7729 in 2019. Through qualitative content analysis \cite{neuendorf2017content}, we observed meaningful shifts in the kinds of content that was shared through tweets in this discourse, moving from straightforward ``naming and shaming'' of companies that used dark patterns to engagement with a broader range of media sources and types of stakeholders---for instance, bringing together designer activism and legal or regulatory guidance.
Finally, a hashtag analysis provides evidence for how the \#darkpatterns discourse metamorphosed from a single conversation around unethical design practices by software engineers and designers to many parallel-yet-intersecting conversations that involve various disciplines, including UX design, privacy, regulation, technology policy, and game design. This pluralism of conversations---joined together through the \#darkpatterns hashtag---sheds light on the multifaceted nature of the discourse around dark patterns.

\begin{table*}[!htb]
 \centering
    \begin{tabularx}{\textwidth}{Xrrrr}
     \toprule
     Year & No. of Tweets & Unique Twitter Accounts & \% Change in Tweets** & \% Change in Twitter Accounts** \\
    \midrule
    \textbf{2010*} & 110 & 95 &  &  \\
    \textbf{2011} & 174 & 151 & 58.18 & 58.95\\
    \textbf{2012} & 239 & 207 & 37.36 & 37.09\\
    \textbf{2013} & 730 & 587 & 205.44 & 183.57\\
    \textbf{2014} & 1091 & 869 & 49.45 & 48.04\\
    \textbf{2015} & 1222 & 895 & 12.01 & 2.99\\
    \textbf{2016} & 2151 & 1288 & 76.02 & 43.91\\
    \textbf{2017} & 2730 & 1570 & 26.92 & 21.89\\
    \textbf{2018} & 6141 & 4409 & 124.95 & 180.83\\
    \textbf{2019} & 7729 & 5425 & 25.86 & 23.4\\
    \textbf{2020} & 5032 & 3449 & -34.89 & -36.42\\
    \textbf{2021*} & 3601 & 2376 &  &  \\
    \bottomrule
    \end{tabularx}
     \caption{Number of tweets and unique Twitter accounts by year. *Beginning on first reference in June 2010 and ending in April 2021. **Year-over-year change by percentage.}
     \label{history}
\end{table*}

\subsubsection{Evolution of Tweet and Tweet Author Volume}

Our analysis of the dark patterns dataset revealed that 30,951 tweets were generated from the first use of the hashtag in 2010 to April 30, 2021, the data collection cut-off date for this study. Of these, 24,261 were tweets and 14,124 were retweets (Figure~\ref{fig:tweetsvsretweets}). The tweets and retweets were generated by 17,993 unique Twitter accounts that participated in the discourse. As shown in Table~\ref{history}, the number of Twitter users that participated in the discourse grew steadily year over year from 2010 (n=95) to 2019 (n=5425), but then dropped in 2020 (n=3449), the last full year of data collected. Similarly, the number of tweets using  \#darkpatterns showed steady growth from the start of the discourse in 2010 (n=110) up until 2019 (n=7729). There was a decrease in the number of tweets in 2020 (n=5032) compared to 2019 (n=7729). The reason for this drop is unknown, but is likely to be due to the COVID-19 pandemic diverting the attention of users to other health-related matters or other economic or social impacts of the global pandemic.
Tweets in our dataset included a total of 39 languages. Similar to the overall patterns of growth in tweets and Twitter accounts, there was a steady growth in the number of languages in the discourse from 2010 (n=3) to (n=31) in 2019, with a drop in 2020 (n=26). While we did not analyze tweets in languages other than English, the large number of languages may be a representative measure of the global reach of the discourse around \#darkpatterns, illustrating how Twitter users globally are expressing concern about similar phenomena relating to technology manipulation, using \#darkpatterns as a shared entry point. Taken together, these insights about the number of tweets, participants, and languages reveal how the Twitter discourse related to \#darkpatterns has evolved from a small conversation involving a few relatively homogeneous Twitter users in 2010 to a global discourse around technology manipulation as of 2021.

\begin{table*}[!htb]
 \centering
    \begin{tabularx}{\textwidth}{p{.15\textwidth}Xrr}
    \toprule
    \textbf{Year} & \textbf{Title} & \textbf{Engagement Metrics*} & \textbf{Linked Resource Type} \\
    \midrule
    2011 & EU bans pre-ticked website boxes to aid consumers & 36 & Mainstream Media\\
    2012 & Apple Has Quietly Started Tracking iPhone Users Again, And It's Tricky To Opt Out & 78 & Online Media\\
    2013 & Deceptive Design formerly darkpatterns.org & 84 & Practitioner Resource\\
    2014 & Some Dark Patterns now illegal in UK & 75 & Practitioner Blog\\
    2015 & Using Open Experience Design and Social Networking to Stamp Out Dark UX & 42 & Practitioner Blog\\
    2016 & \#DarkPattern spotted: graying out active buttons to trick users into writing reviews & 189 & Twitter User\\
    2017 & 3 Chilling UX Dark Patterns Businesses Need To Kill Right Now & 341 & Practitioner Blog \\
    2018 & How Dark Patterns Trick You Online & 471 & Video Sharing\\
    2019 & An Image of depicting dark patterns on ecommerce platform & 1500 & Twitter User\\
    2020 & Bookmark Manager: A Cognitive Bias, Mental Model or Dark Pattern explained every time you open a new tab to help improve your decision-making process. & 185 & Digital Tool\\
    2021 & Bringing Dark Patterns to Light Workshop & 106 & Regulatory Agency\\
    \bottomrule
    \end{tabularx}
     \caption{Most frequently shared resource by year. *This metric includes a combination of the frequency of shares, likes, and retweets with these resources.}
     \label{mediasource}
\end{table*}

\subsubsection{Evolution of Linked Resources}

Through qualitative content analysis, we sought to describe the content of the most linked resource in each year of the discourse. While our goal is not to provide an exhaustive account of all media articles and other resources shared via tweet in our dataset, an analysis of most common linked resources in each year provide an entry point to describe common shared concerns relating to \#darkpatterns in each year of the conversation. As shown in Table~\ref{mediasource}, our analysis revealed that users participating in the \#darkpatterns discourse relied on diverse sources to support their campaign against technology manipulation, including news articles from major news media organizations, blog posts from technology practitioners, YouTube videos, regulatory workshops, and participation in conferences, among other sources.

These linked resources shown in Table~\ref{mediasource} provide a skeleton for a brief historical overview of how the concept of dark patterns has been interpreted in relation to other important technology, regulatory, and historical events which have impacted the evolution of this specific hashtag, building on our historical overview of dark patterns in Section~\ref{sec:darkpatternsbackground}. In 2011, the most-shared media article was a report about the European Union's ban on the use of pre-ticked boxes on websites. This ban mandated shopping websites to provide users with an option of explicitly opting in or out of additional services on their platform. The article cited an example of the forced inclusion of travel insurance when buying airline tickets as an instance of this manipulation. Similarly, the most shared media article in 2012 was the naming and shaming of Apple for tracking iPhone users and making it difficult for them to opt out.  This news article was first tweeted by one of the participants in the discourse and then retweeted by several other Twitter users to spread awareness of a new deceptive practice employed by Apple to collect data from users without their consent. Taken together, these findings reveals the kinds of concerns about dark patterns that Twitter users expressed in the first three years of the discourse, typically with a focus on specific products or companies.

In 2013, the most shared resource was the darkpatterns.org website \cite{brignull_2010}. This website, although introduced in 2010 by Brignull, became dominant in the discourse in 2013 and provided Twitter users with the opportunity to submit reports about their encounters with digital products that used dark patterns techniques. This website also provided information on different types of dark patterns to sensitize users to the diverse range of deceptive techniques---types that grew over time and were eventually augmented by other practitioner and academic sources.
This website aimed to serve as a central resource for supporting a discourse around dark patterns, and several users relied on this resource to enable their participation in naming or combatting  dark patterns. For example, one participant sharing the link to \url{darkpatterns.org} commented: ``\textit{So satisfied to submit a \#Darkpattern.}'' Another user also sharing the resource commented ``\textit{Funny! Watch out the offenders of `Roach motel\ldots'}.'' In 2014, the most shared article was an interview that discussed Consumer Protection Regulations in the United Kingdom. Although the interview shared in this resource framed the amendment as a direct win for dark patterns campaigners, the wording of the regulation did not specifically target organizations using dark patterns or result in the resolution of dark patterns in general.

From 2015 onward, concerns within the discourse transitioned from only naming and shaming companies that used dark patterns to also exploring ways to potentially ``inoculate'' or protect members of the public from the impact of dark patterns through education, while also pointing out advanced forms of dark patterns. For example, the most shared article in 2015 was written by a practitioner who proposed using open forums for ``\textit{Design Experience and Social Networking}'' to combat the influence of dark patterns. This article took an activist stance and served the purpose of reinvigorating users to fight back against companies and designers employing dark patterns. The articles in 2016 and 2017 included advanced examples of dark patterns that users should be able to recognize and avoid.

Similarly, building on the theme of inoculating of users against the harms of dark patterns, the most shared content in 2018 was a YouTube video \cite{darkpatternsyoutube} that was created to sensitize users on how they are being manipulated using dark patterns on several digital platforms. This video has been watched over 1,596,248 times on YouTube as at July 2022, with several of the commenters on YouTube sharing their experiences with dark patterns and technology manipulation in the comment section of the video. In 2019, the most shared media content was an example of dark patterns in an e-commerce platform, while in 2020, another practitioner introduced a browser bookmark that aims to explain to users the type of dark patterns on the web page they are viewing and how to make decisions to protect themselves from falling prey to deceptive design practices.

Finally, in the first third of 2021, the most shared URL was an invitation to the US Federal Trade Commission workshop on dark patterns. This linked resource built upon interest by regulators in dark pattern that was present as a minority discourse through \#darkpatterns since 2018---largely due to the adoption of GDPR as a legal framework for data protection---but for the first time, the regulatory conversation reached critical mass.
The presence of this linked resource, and its shift from a focus on designers or end users to legal remedies (as foreshadowed in the most commonly shared resource from 2011) illustrates the slow and incremental nature of regulation around dark patterns, as governments and states have attempted to regulate or otherwise curtail these practices over time. Notably, the size of engagement and the enthusiasm created around the workshop illustrates a yearning for a stronger government regulation around dark patterns that we have been able to identify since the beginning of the dark patterns discourse.

\begin{figure}
    \centering
    \includegraphics[width=\textwidth]{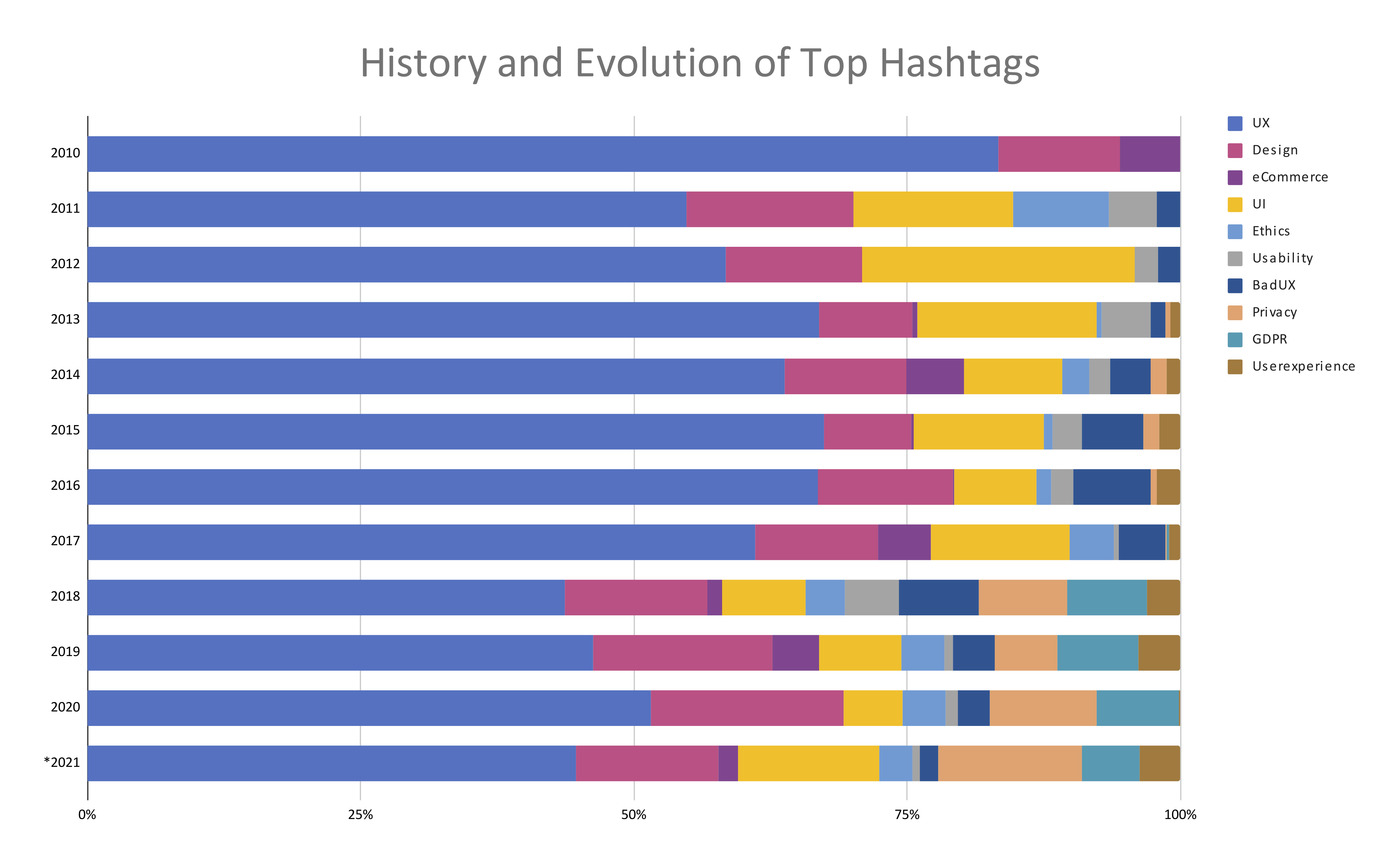}
    \caption{A hashtag analysis of related tags in the discourse from June 2010 to April 2021.}
    \label{fig:hashtaganalysis}
\end{figure}

\subsubsection{Evolution of Co-Occurring Hashtags}
To further illustrate the evolution of the discourse, we analyzed the co-occurring hashtags used in the discourse to better understand how participants sought to frame or connect their interest in dark patterns with other areas of context or concern (Figure~\ref{fig:hashtaganalysis}). Findings from our analysis revealed that a total of 64,087 hashtags were connected to the 27,296 tweets generated in this discourse (SD= 6.3, Mean=2.35). Prior research ~\cite{rho2019hashtag} has shown that online movements employ hashtags for various objectives, including to improve content discoverability, define their discourse, and as a means of connecting related issues to the conversation. We employed a similar lens and identified the top 10 technology-related hashtags that were employed across each year of the discourse to investigate the different fields, concepts, and ideas that participants sought to connect to the \#darkpatterns discourse and how those hashtags demonstrated evolution of the discourse over time.

Findings from our hashtag analysis revealed that the top ten technology-related hashtags which co-occurred with \#darkpatterns during multiple years of the discourse were: \#eCommerce (n=706), \#privacy (n=1603), \#gdpr (n=1281), \#ux (n=14880), \#design (n=3831), \#ui (n=2569), \#ethics (n=911), \#usability (n=354), \#badux (n=1156), and \#userexperience (n=805). This range of hashtags illustrates the plurality of lenses through which concerns about dark patterns has been expressed, including ethical, contextual, regulatory, and professional lenses. A review of the findings shows that the prominent hashtags in 2010---the first year of the discourse---were about e-commerce, UX, and design and how these contexts framed the use and problematic nature of dark patterns. This early combination of hashtags suggests that the primary concern during the early phase of the \#darkpatterns discourse was about the use of dark patterns in eCommerce and shopping websites, demonstrating economic harms. This early understanding of dark patterns was also echoed in the use of e-commerce settings in several dark pattern types proposed by Brignull, including ``sneak into basket,'' ``price comparison prevention,'' and ``hidden costs'' \cite{brignull_2010}. In 2011, additional co-occurring hashtags including \#usability, \#badux, and \#userexperience were introduced to the conversation. The introduction of \#badux demonstrated the interest of participants in the discourse in describing what ``good UX'' should look like, thereby expressing that \#darkpatterns are a normative deviation from those expectations. Similarly, \#ethics was introduced into the conversation in 2011, cementing an explicitly normative position against \#darkpatterns. Additionally, \#privacy was introduced into the discourse in 2013, gaining prominence within the discourse by 2015. \#privacy became one of the most dominant hashtags used in the discourse from 2018 onward, representing a meta-level shift from a description of dark patterns in specific contexts such as e-commerce to broader concerns regarding how dark patterns might impact user privacy or data protection. Additionally, following the introduction of GDPR in 2016 and its formal implementation in 2018, the \#gdpr hashtag also became a consistently employed hashtag starting in 2018.

Examining the content of the tweets associated with combinations of hashtags---for example, tweets utilizing both \#privacy and \#darkpatterns---our analysis revealed that participants often employed \#privacy and \#darkpatterns together when they experienced privacy-related instances of dark patterns or when engaging in a conversation at the intersection of dark patterns and privacy. For example, one participant described their experience with ``unreadable'' terms and conditions documents as examples of dark patterns with ``\textit{malicious \#privacy settings}'' that often ``\textit{compromise user's freedom.}'' While another user shared a report about privacy: ``\textit{Study to illustrate privacy threatened by big tech companies as published by Norwegian Consumer Council, showing examples of misleading privacy policies.}'' Other users also employed the \#privacy in combination with \#darkpatterns to notify companies of the implications of privacy related dark patterns, with a Twitter user sending a signal to other culprit organizations and commenting: ``\textit{Another drop in the bucket! Privacy lawsuit on LinkedIn privacy}.''

Other participants directly characterized dark patterns as a manifestation of BadUX. For example, a Twitter user while sharing their experience with LinkedIn Newsletter reproached them saying: ``\textit{\#badux. Unsubscribing to your newsletter is not working, \@LinkedIn}.'' Other participants characterize dark patterns as a form of unethical design, as it takes advantage of users, with one Twitter user commenting that ``\textit{Increasing a hook to products and decreasing our happiness is a \#darkpattern, encouraging endless consumption}.'' Continuing with the theme of ethics and hinting about the need for regulation, another Twitter user expressed their disapproval of dark pattern and commented: ``\textit{I cannot makes sense why \#tech companies are 'authorized' to trick us into browsing, buying, clicking, and so on. They have our personal \#data. Isn't it the case that \#consumerlaw should be more stricter \#online?}.'' Another Twitter user encouraged their personal network to fight against this unethical design practice commented: ``\textit{Things will be better if more and more of us stand up to this}''---framing the fight against dark pattern as a normative battle between members of the public and technology practitioners. Similarly, the use of the \#gdpr in combination with \#darkpatterns had multiple use cases. Some users employed the \#gdpr as means of highlighting that help has come. Others employed the tags to notify companies using dark patterns of the existence of the policy against such practices, with one user explicitly calling out Mashable: ``\textit{How about being \#GDPR compliant, @mashable, rather than trying to fuck up your readers?}'' Other Twitter users also employed the GDPR hashtag to express that the regulation is not doing enough to curtail instances of user manipulation, with one user commenting that ``\textit{A new turn of \#Darkpatterns with the rise of The whole \#GDPR thing on the \#web, with things like: `this site will suck without cookies.'}'' Furthermore, policymakers often used the GDPR hashtag as a means of creating awareness of the work they were doing around dark patterns with one regulator tweeting ``\textit{Our collection of reports against Google's deceptive practices \& design, videos, and our \#GDPR complaint.}'' Altogether, these findings surface the multiple ways users employed hashtags to enrich the Twitter discourse related to dark patterns from different contexts and disciplinary lenses.

 \begin{table*}[!htb]
    \begin{tabularx}{\textwidth}{lXXXXXp{.125\textwidth}p{.125\textwidth}}
       \toprule
       \textbf{Years} & \textbf{\#eCommerce \& \#darkpatterns} & \textbf{\#privacy \& \#darkpatterns} & \textbf{\#ethics \& \#darkpatterns} & \textbf{\#gdpr \& \#darkpatterns}\\
       \midrule
        \textbf{2010 - 2012} & We will not use dark patterns & Bad UX practices aka dark patterns explained & Gamification is an application of dark patterns?  &  ---\\
        \textbf{2013 - 2015} & This Amazon dark patterns makes my web spidey sense go bubbly & LinkedIn privacy lawsuit is a just a drop in the bucket & Design, White Lies, and Ethics & ---\\
        \textbf{2016 - 2018} & Avoid these 13 examples of dark patterns in ecommerce & Example of Hostile Web Privacy & The more of us who stand up to this the better things will be & See an EULA/Privacy Agreement for software to help with GDPR compliance\\
        \textbf{2019 - 2021} & Study of over 11,000 online stores finds dark patterns on 1,254 websites & Really enjoying this talk about the influence of design on privacy & I cannot understand why tech companies are authorized to trick us into buying, subscribing etc. & GDPR had created a new wave of dark patterns pushing a cookie choice.\\
      \bottomrule
     \end{tabularx}
     \caption{An illustration of tweets associated with the \#darkpatterns and other technology-related hashtags}
     \label{evolution}
 \end{table*}

\subsection{RQ2: In what ways has the composition of Twitter users in the \#darkpatterns discourse changed overtime?}
In this section, we present the results of the evolution of composition of Twitter users in the \#darkpatterns discourse from 2010 to 2021 using two approaches. First, we employed a \textit{network analysis} to investigate and visualize the evolution of dominant professional fields represented in the discourse. Second, we conducted a \textit{qualitative content analysis} of the user description of Twitter accounts from our dataset to identify the ways Twitter users in the discourse described themselves.

\subsubsection{Evolution of Professional Fields in the \#darkpatterns Discourse} \label{networkanalysis}
\begin{figure}
    \centering
    \includegraphics[width=\textwidth]{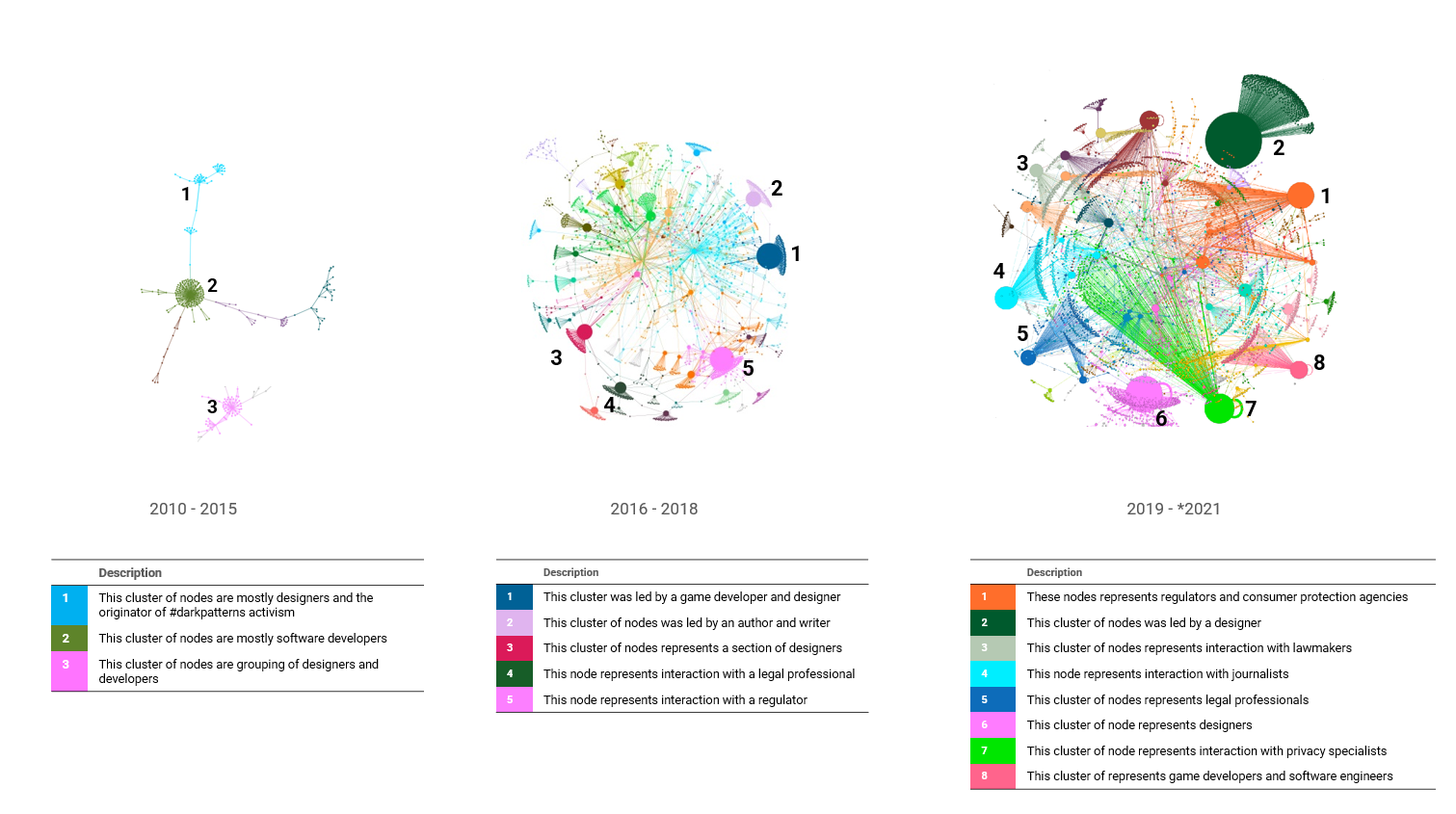}
    \caption{A network analysis of the discourse from June 2010 to April 2021, divided into phases.}
    \label{fig:networkanalysisphases}
\end{figure}

Results of our network analysis, as shown in Figure ~\ref{fig:networkanalysisphases}, revealed a continuous expansion in the number and diversity of professional fields represented in the dark patterns discourse over 10 year period. As described in Section \ref{networkanalysis}, we constructed the network graphs by using Twitter accounts as nodes and retweets, replies and quoted tweets as the edge connecting two accounts. Hence, large-sized nodes (circles) like the ones represented on our network graphs Fig \ref{fig:networkanalysisphases} implies that more Twitter users interacted with the content of that account through retweets, replies, and quoted tweets. Our interpretivist analysis focused mainly on the interactions around these large-sized nodes since they represented Twitter accounts whose content was most interacted with.
Results from our analysis shows three main phases of evolution, including Phase 1 from year 2010 to 2015 (representing 1388 accounts), Phase 2 from year 2016 to 2018 (representing 5042 accounts), and Phase 3 from year 2019 to 2021 (representing 8000 accounts).

\paragraph{Phase 1 (2010 to 2015): The Era of Intra-Disciplinary Discourse}

Results from this phase of our network analysis revealed that design practitioners and software developers were the most dominant professional fields represented during the 2010 to 2015 phase of the discourse (Figure ~\ref{fig:networkanalysisphases}, left). This network graph captured 3,566 tweets across 1,388 unique Twitter accounts. Our network graph shows a tight cluster of nodes connecting to the founder of the \#darkpatterns movement, @harrybr (blue node \#1) and then breaking off in multiple directions, signifying that software developers (\#2) and designers (\#3) were mostly interacting with the content from Brignull and breaking off into groups to discuss what dark patterns meant for them.
A further analysis of the nodes in the graph reveals that other than these two fields, there were no obviously dominant accounts within this phase of the discourse. Instead, the network shows that multiple nodes were relatively similar in size, thus indicating a relatively comparable degree and pattern of engagement with their content. Additionally, the fact that a majority of the cluster of nodes consists of designers and software developers also suggests a limited scope to the discourse around technology manipulation at this time. This interpretation aligns with results from our hashtag analysis which similarly revealed a limited and highly focused hashtag usage in the early phase of the discourse. Interestingly, the network graph for this phase revealed that the professional fields most directly accountable for the implementation of dark patterns were also the ones engaging in the Twitter discourse about technology manipulation.

\paragraph{Phase 2 (2016 to 2018): The Rise of Legal and Regulatory Interest}

The network graph for this phase of the discourse revealed an increase in the diversity of professional fields that became dominant within the \#darkpatterns discourse. This network graph captured 11,022 tweets across 5,042 unique Twitter accounts. Our analysis of top nodes in the network graph (Figure~\ref{fig:networkanalysisphases}, center) reveals that legal practitioners (\#4), regulators (\#5), authors/writers (\#2) became dominant in the discourse, in addition to the continued participation of designers (\#3) and software developers (\#1) from the prior phase. A visual analysis of the network graph reveals an outward explosion of different dominant clusters of nodes and illustrates an evolution of the \#darkpatterns discourse from a focused conversation around design malpractice to a more complex discourse that transcends design and software development to include the potential legal implication of dark patterns and the need to protect consumers from the impact of these practices. This is evidenced by the increasing prominence of regulators and legal professionals within the discourse during this phase.
Furthermore, this phase of the network analysis coincided with the announcement of GDPR in 2016 and the commencement of its implementation in 2018. At this point, the discourse around dark patterns was no longer primarily about deceptive interface design but, also about the downstream implications of those deceptive practices on members of the public and legal implications for businesses implementing such practices.

\paragraph{Phase 3 (2019 to 2021): Concretization of a Trans-disciplinary Discourse}
Results from our network graph for this phase shows a continued increase in the number and diversity of professional fields that had prominence within the discourse. This network graph captured 16,362 tweets across 8,000 unique Twitter accounts. Our graph revealed that federal lawmakers, privacy-oriented companies, regulators/government agencies, academic researchers, consumer protection agencies, lawyers, and journalists also contributed to dominant nodes in this phase of the discourse as can be seen from the (Figure ~\ref{fig:networkanalysisphases}, right) which shows the different cluster of nodes. This increase in the diversity of professional fields participating in the discourse parallels the introduction of GDPR and could have induced several professional fields to join the discourse in order to make sense of what the regulation means for them. Furthermore, our analysis of the network graph reveals that there were stronger edges than during the prior phases, revealing increased engagement in the discourse around dark patterns from different disciplinary lenses.
Altogether, it is interesting to note that the increased dominance of multiple fields and the flavors of conversation they introduced were all taking place in parallel with the continued naming and shaming of companies using dark patterns by everyday users, and all conversations were solidified under the umbrella of dark patterns---evidence of the emergence of dark patterns as a transdisciplinary concept.

\subsubsection{Users Descriptive Analysis}
\begin{figure}
    \centering
    \includegraphics[width=\textwidth]{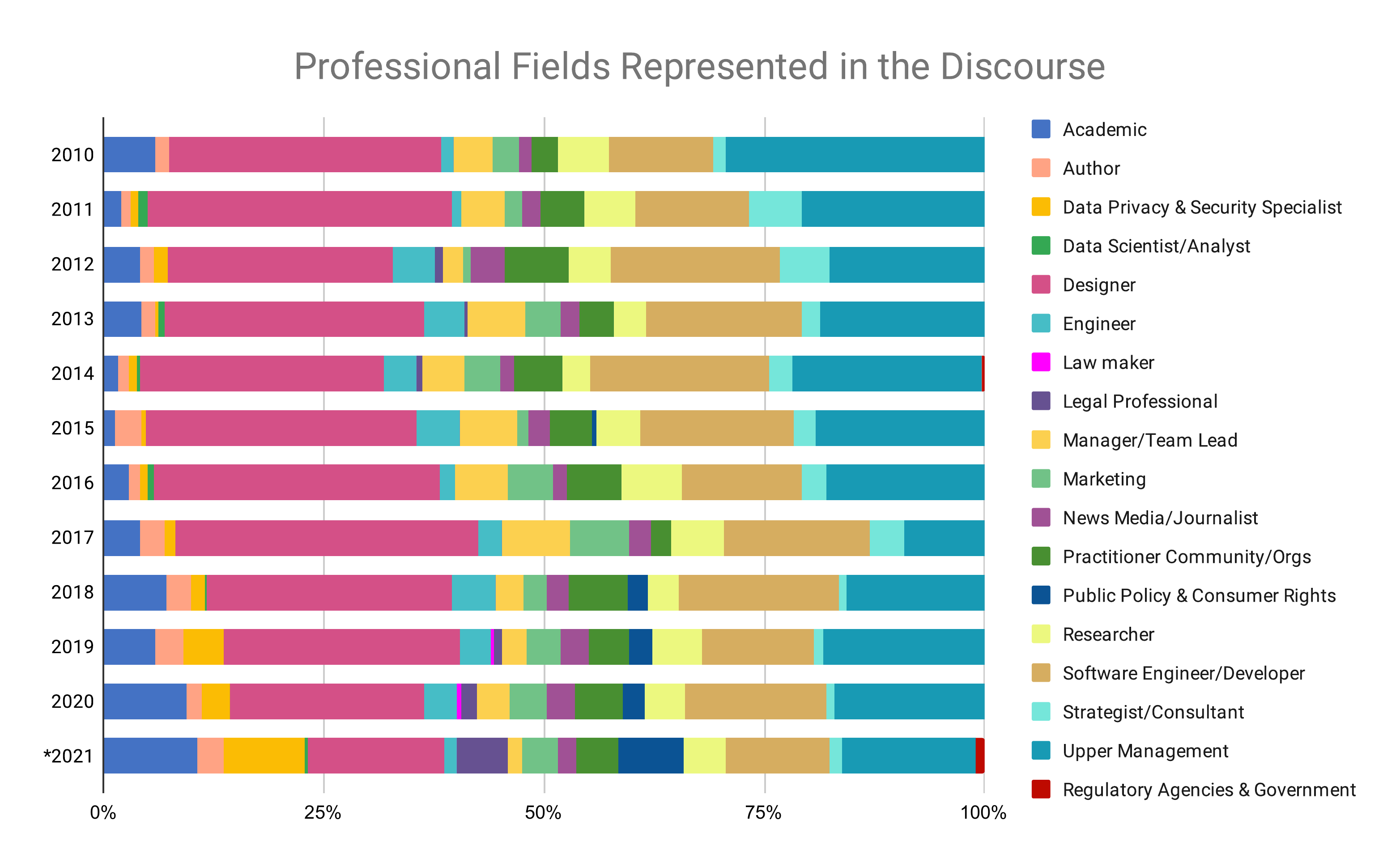}
    \caption{An analysis of the evolution of professional fields represented in the discourse from June 2010 to April 2021.}
    \label{fig:userroles}
\end{figure}

After completing our analysis of the interaction between the professional fields represented in the discourse from a macro lens using network analysis, we transitioned to conducting a qualitative analysis of user descriptions of Twitter accounts that participated in the conversation. Our goal for this analysis was to examine more closely, the affiliations of participants in the discourse and the ways in which these affiliations evolved as the discourse around dark patterns progressed. We conducted this investigation by analyzing open codes from the content analysis of the selected 5,830 tweets. Each Twitter account was coded based on the profile description provided by the Twitter users themselves. For example, one Twitter account disclosed: ``\textit{I am a UX Designer working on healthcare products}.'' As a limitation of this study, user descriptions were collected at the time of data capture and do not necessarily reflect the profile description at the time the tweet was made. Findings from this analysis revealed that there were 733 such distinct professions or roles based on the profile description of the Twitter users based on our initial open codes. A further categorization of these roles according to professional groupings (e.g., combining User Experience Designer and UI Designer to Designer) resulted in 18 distinct categories of professions that retained the diversity implied by the initial roles. Fig~\ref{fig:userroles} illustrates this finding.

An analysis of the these categories revealed that participants in the first year of the discourse---which was 2010---were mostly designers, researchers, and software developers. During 2011, our analysis revealed the entrance of Data Privacy \& Security Specialists into the discourse, however, it took until around 2016 for this group to become dominant in the discourse. Furthermore, regulatory agencies, public policy, and consumer rights advocates joined the discussion in 2014 and 2015, while federal lawmakers became prominent in 2019. Unlike the network graphs, results from this analysis revealed that marketers, journalists/news media organizations, members of the academic community, practitioner groups, and researchers, were constant participants throughout the duration of the discourse, although interaction with their content were much smaller when compared to designers and engineers, hence resulting in poor visibility on the network graph. Notably, this finding reveals that there were several other discourses taking place in the background, including several other fields contributing to the discourse in their own ways that were more idiosyncratic or connective in ways that were not documented on the network graph. Some fields went through the phase of emergence and eventually evolved to prominence within the general composition of the dark patterns discourse, while others continued as a background conversation throughout the discourse.

\subsection{RQ3: Who do participants in the discourse think they are talking to and how has this evolved over the years?}

\begin{figure}
    \centering
    \includegraphics[width=\textwidth]{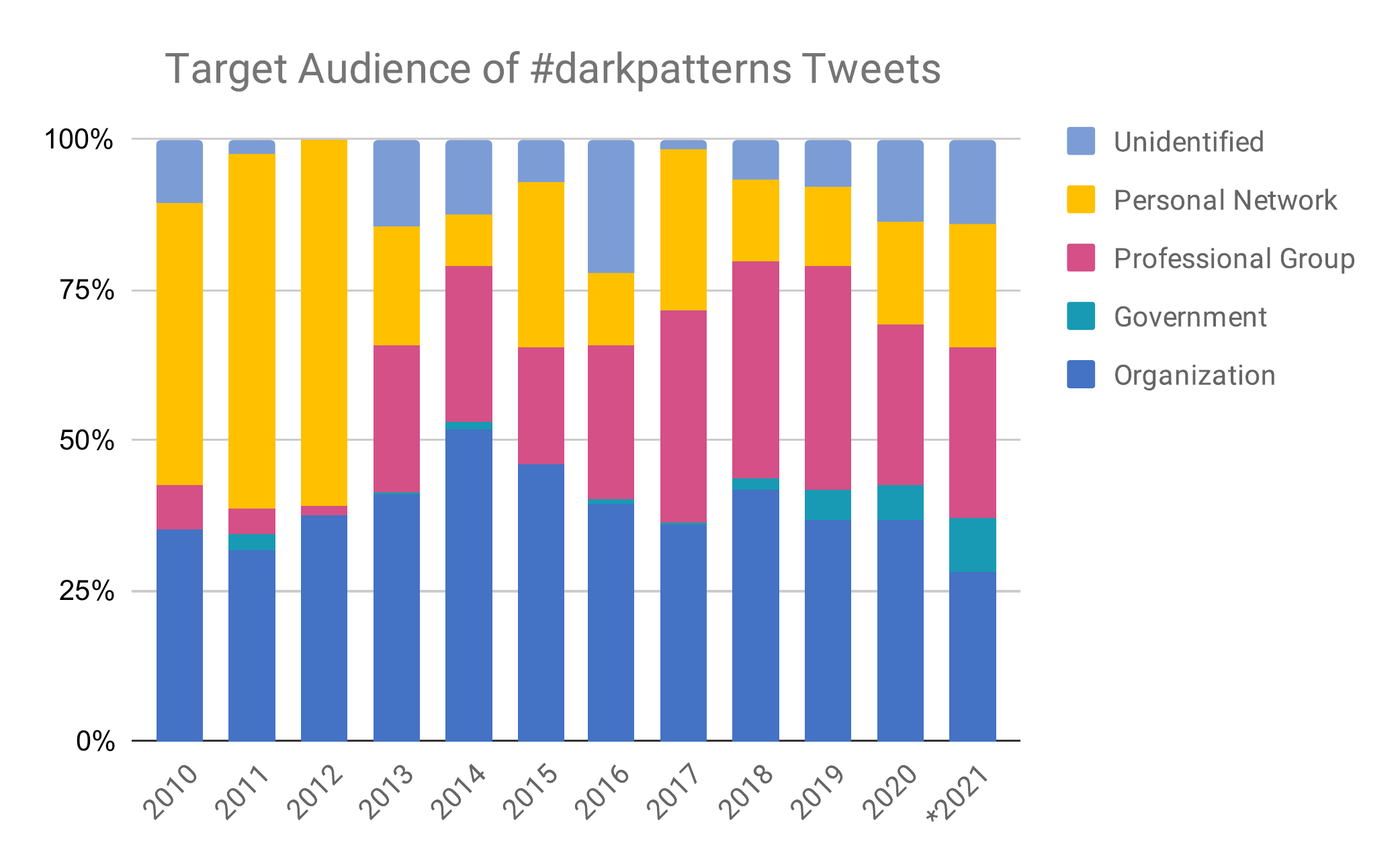}
    \caption{Who the users thought the intended audience of their tweets were from June 2010 to April 2021}
    \label{fig:directionevolution}
\end{figure}

We employed Marwick and Boyd ~\cite{marwick2011tweet}'s lens which describes how Twitter ``\textit{users imagine the audience evoked through their tweets}'' to indicate the target audience. We conducted a content analysis of the 5,830 English-language tweets from our dataset to investigate the intended audience for the \#darkpatterns Twitter discourse and the ways in which this perceived audience evolved over time. We engaged in a reflexive thematic analysis by identifying tweets that illustrated the story of evolution during the previous stages of coding.
We identified four intended audiences for the tweets: government, private organizations, professional groups, and their personal network. We also identified four different objectives that foregrounded the focus of their tweet in relation to this perceived audience, which included: \textit{sense-making} to interpret dark patterns, \textit{activism} to fight against dark patterns, \textit{sharing} their experiences with dark patterns, and \textit{foregrounding} regulatory implications or non-compliance.

\subsubsection{Intended Audience of \#darkpatterns Tweets} Tweets in the \#darkpatterns discourse were directed towards four main types of intended audiences, including: 1) Government and entities who are expected to hold regulatory or law-based control over dark patterns (n=165/5830); 2) Private Organizations utilizing dark patterns in their products and services (n=2191/5830); 3) Professional Groups that the user belongs to or are calling out through the tweet, such as UX designers, software engineers, researchers, lawyers, or consumer protection agencies (n=1600/58300); and 4) Personal Networks of the user (n=1199/5830). A total of 225 tweets were coded as unidentified with respect to the intended audience of the tweets, and all other tweets were coded non-exclusively.

An analysis of the evolution of the intended audience of tweets, presented in Figure \ref{fig:directionevolution}, revealed changes in the proportions of various intended audiences over time. In the early phase of the discourse from 2010 to 2012, most of the tweets were intended for the user's personal network. For example, one user posted a resource about dark patterns for their personal network mentioning: ``\textit{Find out more what \#darkpatterns are in design.}'' Another user shared a tweet about their experience with dark patterns in an airlines context with their personal network, saying: ``\textit{Many Back end, not directly UI, dark patterns on budget airline sites.}''
From 2013 to 2016, users directed a substantial number of their tweets towards private organizations implementing dark patterns. For example, a user shared their experience with dark patterns they faced by calling out companies in a tweet: ``\textit{Twice this week! Screwed by \#darkpatterns. Thanks to @ebay \& @amazon}.'' A similar tweet shared: ``\textit{My whole network got spammed last week, without permission. Really unhappy with @AngelList.}'' In these cases, the users expressed their frustration by explicitly calling out the companies using dark patterns in their user interfaces. From 2018 onward, the number of tweets directed towards the government began to increase. Some of these tweets related to users calling for the attention of government agencies relating to dark pattern infractions by different organizations. For instance, a tweet calling out on Facebook and violation of GDPR stated: ``\textit{\#GDPR slap a huge fine with \#DarkPatterns taken to another new level. @facebook on @android asks for SMS and Contacts access without justification.}'' This tweet is also an example of how users began to use hashtags in combination with \#darkpatterns to call attention from the government. Other tweets directed towards the government were posted to provide feedback and recommendations about government regulations around dark patterns such as: ``\textit{Legislation must interlink \#DarkPatterns and \#DataPrivacy issues as they go hand in hand, to regulate both of these potential threats.}'' From 2017 and 2021, the discourse was also directed towards private organizations and professional groups (in almost similar patterns) responsible for implementing dark patterns, suggesting that the participants by this stage of the conversation have realized that the responsibility for the implementation of dark pattern resides at both the individual practitioner and organizational levels, with a variety of ecological factors influencing the relationship between these entities.

\subsubsection{Intended Purpose of \#darkpatterns Tweets}
In order to tell a holistic story about the evolution of the \#darkpatterns discourse, we carefully curated a set of tweets to provide differing examples of how \#darkpatterns tweets were framed in relation to different objectives and intentions that changed over time. While our goal is not to find a \textit{definitive} pattern of evolution through these objectives of tweeting, but the we found the objectives to be framed in relation to: \textit{sense-making} to interpret what dark patterns mean, \textit{activism} to fight dark patterns, \textit{sharing} experiences relating to their daily encounters with dark patterns, and most prominently after 2018, \textit{foregrounding} discussions of regulatory interest and non-compliance.

\paragraph{Sense-making to interpret dark patterns:} users engaged primarily their personal network by questioning, making a note-to-self, or commenting ``out loud'' about their acquisition of knowledge regarding how dark patterns work---part of an ongoing and public conversation that described the Twitter user's patterns of sense-making about the concept of dark patterns. For example, a user introducing dark patterns tweeted: ``\textit{Scammy web design practices have a name now called `Dark Patterns' \#darkpatterns}'', defining dark patterns in their own words as sum of different forms of unethical design practices they faced on digital platforms. Another user questioned an UI choice on an e-commerce website saying: ``\textit{Pure mistake or \#darkpattern? Experience on electrical store site gets confusing: filter by low-high price and then, get most expensive first.}''
These findings align with prior work \cite{paul2010understanding} on how groups typically engage in sense-making activities in ``collaborative discourses'' in order to establish a shared understanding of the problem they are either trying to highlight or solve, including work directly focused on community engagement with unethical design practices \cite{Gray2021-xj}.

\paragraph{Engaging in activism to fight dark patterns:} users directed the \#darkpatterns discourse to engage in activism in two primary ways: 1) directing their tweets towards private organizations by naming and shaming organizations for using dark patterns in their digital products; and 2) directing their tweets towards professional groups by calling out practitioners and spreading awareness among practitioner communities about dark patterns on different digital platforms. In one example, resonant with Brignull's initial goals to ``name and shame'' deceptive design practices, a user directly called out Groupon in a tweet: ``\textit{@Groupon are so horrendous. You trick us into a subscription to emails and ignore when we opt to unsubscribe. \#darkpatterns}.'' Similarly, another tweet regarding the store Asda mentioned: ``\textit{Not sure where to start! Oh @Asda, your online grocery store is woeful. Tricking me for allowing substitutions is a \#darkpattern}.'' Interestingly, some practitioners in the discourse also called out other practitioners that they felt were responsible for implementing dark patterns. For instance, one practitioner tweeted: ``\textit{Do not use \#darkpatterns and take responsibility as a software developer \url{http://t.co/wdgoqz6x8X}}.'' Another user directed their tweets to fellow developers, mentioning: ``\textit{Dear \#gamedev friends, a lot of us as are passionate and care about the art form in our games. We're part of an industry that needs to be profitable as well. But, we need to truly learn how \#darkpatterns work}.'' As a member of a professional community, some practitioners also used \#darkpatterns to organize and spread the news about conferences or presentations to sensitize fellow practitioners on how to avoid dark patterns: ``\textit{Join the bright side if you are a \#darkpatterns bandit designing for the short term. Hear @harrybr at \#jboye13 \url{http://t.co/HFCEIfWRjn}}.''

\paragraph{Sharing (daily) experiences with dark patterns:} users framed their tweets to share their daily experiences with deceptive practices and user interfaces on digital products. For example: ``\textit{A prime \#darkpattern. My friend's parents, intelligent people, got tricked into unwanted \#AmazonPrime. Not good @AmazonUK \#ixd}.'' With this tweet, the user was simultaneously sharing their friend's experience with a dark pattern while also calling out Amazon Prime, who they felt was responsible for implementing these deceptive practices. Similarly, another user shared the experience of their colleague: ``\textit{My flatmate got furious with \#darkpatterns \#usability. She was trying to book tickets with \#Ryanair. A real life user test.}'' By sharing daily experiences and being able to connect deceptive experiences to the concept of \#darkpatterns, users were able to both call out companies and also spread awareness about the topic among their personal networks, representing a more personal and grounded form of activism in the discourse.

\paragraph{Foregrounding regulatory implications or non-compliance}
In the latter portion of the discourse, users began to reference specific regulatory actions and call for both companies and governmental entities to address dark patterns. This objective was illustrated through tweets calling out companies and warning them regarding the potential impact of dark patterns-related sanctions on their business, while also providing feedback to governmental agencies about a range of issues to consider when constructing regulation around dark patterns.
For example, a user posted explicitly to the US FTC (Federal Trade Commission)
in relation to dark patterns saying: ``\textit{Good thread. One thing I hope the @FTC does while digging into \#DarkPatterns is leverage the enormous expertise of its own staff in the Division of Advertising Practices. Many so-called \#DarkPatterns are online analogs of problematic advertising practices.}'' Here, the user provides a suggestion to the FTC that describes how this organization might regulate dark patterns by drawing from their own existing regulatory practices in an advertising context.
As mentioned in Section~\ref{background}, following the beginning of GDPR enforcement in 2018, the majority of regulatory engagement in our dataset related to GDPR and focused on potential organizational non-compliance to the GDPR guidelines. One such tweet is as follows: ``\textit{I am unable to tell which option is “in” or “out”  \#UX.  Great example of \#DarkPattern by @Bloomberg and @TrustArc. How can companies build trust using these techniques? (Even more when you realise there are 96 trackers!!!) \#GDPR. }''

\section{Discussion}
In this paper, we have described the ways in which the \#darkpatterns discourse emerged and evolved over the last ten years on Twitter. Through our analysis, we have highlighted shifts in the professional fields represented in the discourse, the target audience of the tweets, and the intentions of building the discourse over these years, among others. In the following subsections, we first discuss and speculate about the future trajectory of the dark patterns discourse, describing evidence of both epistemological pluralism and of the potential for discourse fracture. Second, we define and link the concept of \textit{socio-technical angst} to our analysis, identifying potential implications of this angst when directed towards technology practitioners and regulators by social media users in response to unethical design practices and technology manipulation.

\subsection{Evidence of Pluralism and Signs of Fracturing through the Evolution of the Discourse}

Leveraging our analysis of evolution of \#darkpatterns over time on Twitter, our findings clearly illustrate the additive nature of the dark patterns discourse. As we have demonstrated, the discourse grew rapidly through increased engagement with the topic of dark patterns, which was concomitant with the expansion of the variety of resources circulated through the discourse, the accumulation of varying hashtags being used alongside \#darkpatterns, and the rapid increase in users overtly expressing their daily struggles and frustrations with dark patterns in digital products. However, our findings indicate that a primary driver for the growth of the discourse was due to a robust increase in both participation and involvement from multiple disciplinary points of view, including designers, software developers, legal bodies, regulatory activists, and academic scholars, among others. This emerging \textit{pluralism} of the discourse revealed positive uptakes for the study of dark patterns,
but also highlights the potential negative side-effects of \textit{fracturing} in the discourse, which we discuss further in this section. When foregrounding the pluralistic nature of the discourse, as supported by our findings, we demonstrate that there is no unitary characterization of the discourse around dark patterns, but rather a range of discourses that are intertwined and connected to the umbrella concept of dark patterns, which also exist as segments in relation to the whole. This pluralistic characteristic of this discourse perhaps provides a new context and point of comparison within the CSCW literature as compared to  social activist contexts that are often framed in more monolithic ways.

The nature of the discourse demonstrates pluralism in a number of ways. For instance, tweets have incorporated a variety of hashtags used alongside \#darkpatterns that called out particular disciplines (such as \#UX), emotions and attributes (such as \#bad, \#evil), or human values (such as \#privacy). Over a decade, the discourse also had a variety of intended audiences in different compositions---from personal networks to share about the concept of dark patterns, professional networks to fight against dark patterns, and even calling for the involvement of legal bodies to discuss and regulate dark patterns in digital products. The introduction of new disciplines into the discourse enriched the entire conversation around dark patterns and provides both the participating Twitter users and dark patterns researchers with a new and improved understanding of the different manifestations of dark patterns as they have unfolded over time. Additionally, findings from the network analysis (Figure~\ref{fig:networkanalysisphases}) revealed that participants from different disciplines (e.g., design, software engineering, law) interacted with each others' content and perspectives on dark patterns that often were derived or supported from distinct knowledge bases. These findings illustrate that various disciplines within the \#darkpatterns discourse may be mutually influencing each other in varying ways that support both a broader umbrella discourse and distinct-yet-related sub-discourses. With the range of these disciplinary orientations engaging in the same discourse, there may be a higher likelihood that these conversations might build or otherwise support translational knowledge to aid the action of lawmakers or legal professionals, or to identify new factors to influence the decisions of designers and software engineers.
As seen in the network analysis for Phases 2 and 3 (Section~\ref{networkanalysis}, center and right) and the evolution of the discourse towards the expansion of engaging multiple-disciplinary views on dark patterns, we also identified a beginning of translucent forms of fracture, primarily along disciplinary lines. The meaning of dark patterns pointed different implications and priorities based on the disciplinary role and set of responsibilities. For instance, for a lawyer, the implications of the dark patterns discourse might be to protect the legal interest of the company from falling short of the regulations around technology manipulation, whereas for a designer or software engineer, the implications focus on building ethically-focused products that users directly interact with, thereby positively impacting their day-to-day interactions with the digital products. This potential for fracture may be led by participants that take on activist roles as a member of their discipline within the discourse, as seen in the form of clusters of activists on the different phases of the network analysis. These experiences differently frame the focus of the discourse by  disciplinary role---which while helping to expand the richness and pluralism in the discourse, may also result in a \textit{fracture} due the lack of resonance or alignment across disciplinary bodies of knowledge, perspectives, or priorities. While the present state of the \#darkpatterns discourse incorporates both variety and the possibility of fracture, there may be the need to
stabilize the discourse with the introduction of additional shared vocabulary to address the rapid expansion of related participants.
The current forms of fragmentation may be acceptable;
however, for this trade-off to be meaningful there would need to be, at minimum, a willingness to explicitly collaborate across varying disciplines towards resolving the proliferation of dark patterns in digital systems and building cooperation around areas of disciplinary overlap.

Based on the tendencies towards pluralism and fracture we have identified through our characterization of the \#darkpatterns discourse, we present three potential (and speculative) trajectories of this discourse, with impacts for design, policy, and activism:
\begin{itemize}
    \item \textit{Continued Epistemological Expansion:} If pluralism continues to characterize and dominate the discourse in a positive direction, facilitating continued expansion and providing multiple perspectives, this trajectory could result in supporting the creation and implementation of new methods to evaluate or detect dark patterns in technological interventions at its earliest stages, connecting design and policy to further curtail dark patterns through regulation, while also \textit{advocating} across disciplinary boundaries to build new transdisciplinary knowledge about dark patterns. This trajectory may focus on a unified, holistic, and balanced treatment of dark patterns with equal uptake and participation across a range of disciplinary perspectives.

    Based on the signs of fracturing we observed, we speculate two potential types of trajectories:
    \item \textit{Macro-fracturing:} Given the emerging concept of deceptive design which may displace or disrupt the use of ``dark patterns'' which would continue to motivate the established discourse we have described,
    we speculate that parallel discourses using different vocabulary such ``deceptive design'' and ``dark patterns'' may emerge. This macro-level fracturing may result in new discourses that either splinter based on disciplinary boundaries, or splinter into distinct sub-groups of transdisciplinary conversation. In either case, macro-fracturing might result in diluting regulatory steps already taken using the term ``dark patterns'' and potentially result in the reinvention of transdisciplinary discussions under the banner of ``deceptive design'' that requires the re-establishment of transdisciplinary alignment that we have observed in the discourse under study.
    \item \textit{Micro-fracturing:} While macro-fracturing refers to instances where entirely new discourses emerge, we have already begun to observe micro-fractures within the \#darkpatterns discourse which relate to particular disciplinary points of view, directionality, context, or other characteristics. As the discourse continues to grow in size and complexity (as illustrated in the shift from Phases 2 to 3 in Figure~\ref{networkanalysis}), the plurality of discourses may readily lead to distinct sub-discourses which fracture---not on totalizing lines that create wholly new discourses---but rather as distinct sub-conversations that are still parseable to others within the larger discourse, but perhaps with varying levels of priority or interest. For instance, in Phase 3 of the current discourse, some designers or technology practitioners may prefer to continue to ``name and shame'' without engaging in the conversations regarding regulation and the legality of particular dark patterns.
\end{itemize}

\subsection{Sociotechnical Angst, Technology Manipulation, and Incremental Regulation}

Grounded in concerns on the material impact of unethical technological systems on the lived experiences of everyday consumers, the discourse around dark patterns indicates an increasing and persistent \textit{socio-technical angst} from everyday users of technology products directed towards lasting change in technology practices. By \textit{socio-technical angst}, we refer to the engagement of a broad range of stakeholders in online discourses in considering the problematic nature of deceptive design practices. This ``angst’’ is concretized through this pluralistic discourse, including multiple dimensions: 1) a discussion of deceptive practices from multiple perspectives that includes social and technical causes, with the attendant harms, 2) an outline of implications to reel in or obviate the excesses of design and technology work, considering both technical realities and the broader ecology in which this technical work is carried out, and 3) identification of existing and future means of regulatory control or other means of amelioration. By describing the \#darkpatterns discourse as a form of socio-technical angst, we refer to a complex constellation of conversations that is carried out in a register that is often characterized by outrage, complacency, or resignation. However, we also wish to frame the teleology of socio-technical angst as pragmatic and solution-oriented in nature---in the case of \#darkpatterns, with participants using the transdisciplinary convergence of multiple perspectives to find translational opportunities across professional spheres of knowledge to address the current realities of deceptive design practices. In other words, socio-technical angst is not about mere catharsis (although that is important!), but rather a catharsis that is grounded in both existing socio-technical complexity \textit{and} the need for a socio-technical solution that may be multi-faceted and involved many different stakeholders.

We will briefly describe some of the key facets of socio-technical angst that were present in the \#darkpatterns discourse and then use these facets to point towards the potential utility of this concept in studying other social movements or forms of technologically-mediated activism. In our study context, instances of socio-technical angst were grounded in---and revealed---through participant interactions and engagement, use of metadata to create relationships among content, expansion of the discourse to include new perspectives, and in the potential future growth of the discourse. First, participant interactions and engagement increased the number of individuals discussing these concepts, creating a safe and supportive environment to recognize the importance of deceptive practices and create norms around ``naming and shaming,'' among other forms of valued interaction. Second, metadata such as hashtags were used to build sub-conversations regarding different facets of the overall phenomenon of dark patterns, allowing participants to indicate contexts and points of relevance that they wished to enrich through their participation. This use of metadata allowed for multiple forms of legitimate participation---through linking of concepts, sharing and agreeing with others' content, or extending content through resources or other engagement. Third, the expansion of the discourse through the entrance of more disciplinary perspectives allowed for new forms of argumentation to emerge and find purchase in the overall landscape of socio-technical complexity. For instance, while early discussions were focused on an intra-disciplinary approach to ``solving'' the problem of dark patterns---encouraging companies and designers to do the right thing---later engagement with regulators or legal scholars increased the options for amelioration to include new laws, policies, regulatory frameworks, and other incentives or disincentives to control company and designer behavior. Finally, this notion of socio-technical angst provides a framework through which potential expansion of the discourse could be broached. While most interactions were centered on the cathartic ``angst,'' these instances of angst could be organized around teleology. Are examples focused on causes, such as designer intentions? corporate incentive structures? disciplinary norms? Or are examples focused on social impacts or means of disruption, such as disproportionate harms, means of control through regulation, or new ways of organizing the creation of digital products in ways that increase user agency and autonomy?

CSCW scholars might find value in using this concept of socio-technical angst to provide further precision when describing both the static state and dynamic interactive opportunities within online social movements. In what ways are elements of the discourse as \textit{content} being used to rhetorically frame or enrich a broader movement? How are metadata used to link or translate concepts? And how are a diverse range of participants contributing to various parts of the overall movement, even if their actions are primarily centered in one place (e.g., catharsis for catharsis' sake; exploring the complexity of causes; anticipating future opportunities for change). Consideration of the pragmatism of engagement is perhaps the most central feature of this framing concept---allowing for engagement in futures that are incremental in nature (e.g., regulation, disciplinary codes) while also maintaining focus on factors that drive the problem as a whole.

\section{Conclusion}
In this paper, we analyzed the emergence and evolution of the \#darkpatterns Twitter discourse from June 2010 to April 2021,
providing insight into how the discourse has changed over time in relation to topics of discussion, types of participants involved, and the directionality and potential audiences involved in the discourse.
We used a range of qualitative and quantitative methods to describe
an increase in tweets and range of linked hashtags over time, alongside an expansion of resource types shared as part of the discourse. In addition, the evolution of the discourse included an expansion of disciplinary perspectives and views with the increase in the network of users engaging with \#darkpatterns, and a range of intended audiences that included governmental entities, private organizations, professional groups, and personal networks across multiple types of intentions that included sense-making, spurring activism against dark patterns, sharing experiences, and foregrounding regulatory implications of such patterns.
Based on the evolution of this discourse, we characterize this conversation as additive and pluralistic, bringing together diverse disciplinary perspectives in ways that preserves disciplinary interests. This discourse is also an instance of \textit{socio-technical angst}, through which participants engage in discussion of both technical realities and social context and impact with a directionality towards amelioration of these issues through a socio-technical approach.

\begin{acks}
This work is funded in part by the National Science Foundation under Grant Number 1909714.
\end{acks}

\bibliographystyle{ACM-Reference-Format}
\bibliography{cleanforarxiv}


\end{document}